\documentclass[12pt, draftclsnofoot, journal, letterpaper, onecolumn]{IEEEtran}

\usepackage{amsfonts}
\usepackage[dvips]{graphicx}
\usepackage{times}
\usepackage{cite}
\usepackage{amsmath}
\usepackage{array}
\usepackage{amssymb}

\usepackage{stfloats}
\usepackage{slashbox}
\usepackage{graphicx}
\usepackage{footnote}
\usepackage{booktabs}
\usepackage{array}

\title{Optimization Framework and Graph-Based Approach for Relay-Assisted Bidirectional OFDMA Cellular Networks}
\author{Yuan~Liu, Meixia~Tao, Bin~Li, and Hui~Shen
\thanks{Y. Liu and M. Tao are with the Department of
Electronic Engineering at Shanghai Jiao Tong University, Shanghai,
200240, P. R. China. Email: \{yuanliu, mxtao\}@sjtu.edu.cn.}
\thanks{B. Li and H. Shen are with the Department of Communication Technology
Research, Huawei Technologies Co. LTD, Shenzhen, P. R. China. Email:
binli@huawei.com, nickshenhui@sina.com.}}

\begin{document}
\maketitle

\vspace{-1.2cm}
\begin{abstract}

This paper considers a relay-assisted bidirectional cellular network
where the base station (BS) communicates with each mobile station
(MS) using OFDMA for both uplink and downlink. The goal is to
improve the overall system performance by exploring the full
potential of the network in various dimensions including user,
subcarrier, relay, and bidirectional traffic. In this work, we first
introduce a novel three-time-slot time-division duplexing (TDD)
transmission protocol. This protocol unifies direct transmission,
one-way relaying and network-coded two-way relaying between the BS
and each MS. Using the proposed three-time-slot TDD protocol, we
then propose an optimization framework for resource allocation to
achieve the following gains: cooperative diversity (via relay
selection), network coding gain (via bidirectional transmission mode
selection), and multiuser diversity (via subcarrier assignment). We
formulate the problem as a combinatorial optimization problem, which
is NP-complete. To make it more tractable, we adopt a graph-based
approach. We first establish the equivalence between the original
problem and a maximum weighted clique problem in graph theory. A
metaheuristic algorithm based on any colony optimization (ACO) is
then employed to find the solution in polynomial time. Simulation
results demonstrate that the proposed protocol together with the ACO
algorithm significantly enhances the system total throughput.

\end{abstract}


\section{Introduction}
\setlength\arraycolsep{2pt}

\subsection{Motivation}
In wireless cellular networks, deploying a set of relay stations
(RSs) between a base station (BS) and mobile stations (MSs) is a
cost-effective approach for improving system performance, such as
coverage extension, power saving and cell-edge throughput
enhancement. These advantages are achieved as relay-assisted
cooperative transmission exploits the inherent broadcast nature of
wireless radio waves and hence provides \emph{cooperative diversity}
\cite{LanemanIT03, LanemanIT04, SendonarisTC03}.

However, due to the half-duplex constraint in practical systems
(i.e., a node cannot receive and transmit simultaneously),
relay-assisted communications suffer from loss in spectral
efficiency. Recently, network coding has demonstrated significant
potential for improving network throughput \cite{AhlswedeIT00}. Its
principle is to allow an intermediate network node to mix the data
received from multiple links for subsequent transmission. Physical
layer network coding, as a means of applying this principle in
wireless relay communications has received increasing attention
\cite{ZhangMobicomm06, RankovJSAC07}. One simple but important
example is two-way relaying, where a pair of nodes exchange
information with the help of a relay node. Compared with the
traditional one-way relaying, the two-way relaying overcomes the
half-duplex problem and
provides an improved spectral efficiency in bidirectional
communication \cite{RankovJSAC07, PopovskiICC07, KimIT08}. It is
thus attractive to utilize \emph{network coding gain} in the form of
two-way relaying for more efficient transmission of downlink and
uplink traffic in a cooperative cellular network.

Orthogonal frequency-division multiplexing (OFDM) is an enabling
physical layer technology for spectrally efficient transmission as
well as user multiplexing in broadband wireless networks. An
intrinsic feature of orthogonal frequency-division multiple-access
(OFDMA) is its capability of exploiting the frequency selectivity
enabled \emph{multiuser diversity}. A deep faded subcarrier for one
MS may be favored by another MS. Yet, it is a nontrivial task to
perform subcarrier assignment in an OFDMA system.

The goal of this work is to investigate the aforementioned three
types of gains, namely, cooperative diversity gain, network coding
gain, and multiuser diversity gain, in a relay-assisted
bidirectional OFDMA cellular network. To this end, we present in
this paper an optimization framework for resource allocation and
further propose an efficient graph-based algorithm to utilize these
gains simultaneously.

There are three questions to be addressed in this paper.
First of all, it is known that relaying is not always necessary in
relay-assisted communications. For example, when the channel
condition of direct link is better than that of the cooperative
link, direct transmission will be preferred. Furthermore,
even if relaying is necessary, two-way relaying may not be always
applicable. For example, when the downlink channel is good but the
uplink channel is poor, then only uplink transmission needs relay
assistance, and hence there is no opportunity to employ two-way
relaying. Therefore, the first question we will address is how to
design a unified transmission protocol which can support direct
transmission, one-way relaying and two-way relaying. The second
question to address is how to determine the transmission mode
(direct transmission, one- and two-way relaying) of the  downlink
and uplink traffic for each MS.
This is essentially a problem of opportunistic relaying with or
without network coding.
Thirdly, how to efficiently allocate the subcarriers and select the
RSs is crucial so as to maximize the system total throughput.
The answers to these questions are given \textit{globally} and
\textit{systematically} in this paper.

\subsection{Related Work}

Optimization in both cooperative networks and OFDMA cellular
networks has been extensively studied in the literature (e.g.,
\cite{Song1TW05, Song2TW05, TangJSAC 07, KimTW08, TaoTW08}).
However, only a few attempts have been made very recently to study
the optimization of bidirectional cooperative OFDMA-based cellular
networks \cite{YuJSAC07, LiangVT09, XuINFOCOM09}. Authors in
\cite{Yu JSAC07} present a framework for joint optimization of relay
selection, relay strategy selection, power and subcarrier allocation
in which, however, only conventional one-way relaying is used. By
Lagrange dual decomposition method, the joint optimization problem
is decomposed into per-subcarrier subproblems that can be solved
independently. In \cite{LiangVT09}, authors propose a hierarchical
protocol for one- and two-way relaying in a two-time-slot
time-division duplexing (TDD) mode. In this protocol, the
transmission mode of each MS as well as its assisting RS (if relay
mode is selected) are pre-fixed, and the downlink and uplink
transmission modes for each MS are the same. Then, only joint power
and subcarrier allocation is considered and solved by Lagrange dual
decomposition method as in \cite{YuJSAC07}. Authors in
\cite{XuINFOCOM09} propose
an XOR-assisted cooperative diversity scheme and present a
heuristic algorithm for joint optimization of relay selection,
transmission mode selection, power and subcarrier allocation. This
system operates in frequency-division duplexing (FDD) mode with
fixed sets of data subcarriers and relay subcarriers. The work,
however, does not consider the pairing issue when applying network
coding to combine downlink and uplink traffic, the necessity of
which will be detailed in Section II.

\subsection{Contributions}

In this paper, we consider an OFDMA-based wireless cellular network
that maintains bidirectional downlink and uplink traffic for each
MS. Centralized processing is assumed so that the base station
controls the behavior of all users and relays.
%
%
The main contributions of this paper are summarized as follows:

\begin{itemize}
\item A novel three-time-slot TDD transmission protocol for supporting direct transmission, one- and
two-way relaying in bidirectional cooperative cellular networks is
proposed. In this protocol, each frame is divided into three time
slots. In the first two time slots, BS and MSs transmit the downlink
and uplink traffic, respectively, while RSs remain silent. In the
third time slot, RSs help to forward the downlink and uplink traffic
only when necessary.
\end{itemize}

\begin{itemize}
\item Using the proposed three-time-slot TDD protocol,
we formulate a joint optimization of bidirectional transmission mode
selection, subcarrier assignment, and relay selection for maximizing
the system total throughput. For simplicity, uniform power
allocation is considered.
There are three main distinct features about our problem
formulation. First, we develop five feasible transmission modes,
instead of three (direct transmit, one- and two-way relaying) for
the bidirectional traffic to select. Second, for each MS, the uplink
and downlink traffic always occurs in pair so that we can exploit
the network coding gain through two-way relaying as large as
possible. Third, each traffic pair can contain multiple parallel
sessions, each of which can be assigned a different transmission
mode and take place on a different set of subcarriers.
%
\end{itemize}

\begin{itemize}
\item The joint optimization problem is a combinatorial problem and NP-complete.
To make it more tractable, we adopt a graph theoretical approach.
First, we establish the equivalence between the original joint
optimization problem and a maximum weighted clique problem (MWCP) in
classical graph theory. A metaheuristic algorithm based on any
colony optimization (ACO) is then employed to solve the MWCP problem
in polynomial time.
\end{itemize}
%

\subsection{Organization}

The remainder of this paper is organized as follows. Section II
introduces the system model and the proposed transmission protocol.
Section III presents the optimization framework that jointly
considers subcarrier allocation, transmission mode selection, and
relay selection. Section IV presents an efficient algorithm to solve
the optimization problem by a graphic approach. Section V provides
extensive simulations to verify the effectiveness of the algorithm.
Finally, we conclude the paper in Section VI.

\section{System Model and Proposed Transmission Protocol}

We consider a single cell OFDMA wireless network with one BS,
multiple MSs, and multiple RSs. Each MS can communicate with the BS
directly or through one or multiple RSs. The communication is
bidirectional and subject to the half-duplex constraint.
%

\begin{figure}[tbhp]
\begin{centering}
\includegraphics[scale=.8]{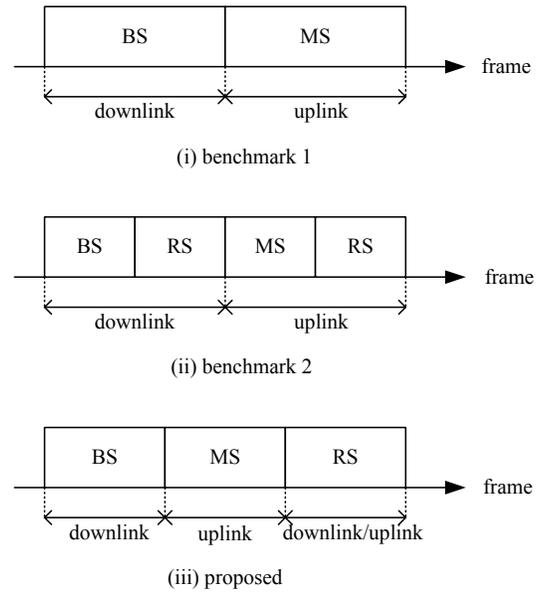}
\vspace{-0.1cm}
 \caption{Three bidirectional transmission schemes.}\label{fig:benchmarks}
\end{centering}
\vspace{-0.3cm}
\end{figure}

In traditional cellular networks where no relay is used, to support
both downlink and uplink transmission, either TDD or FDD has to be
applied. In particular, in TDD system, as shown in
Fig.~\ref{fig:benchmarks}(i), the transmission frame is divided into
a downlink subframe and an uplink subframe, both on the same
frequency band but in two different time slots. When RSs are
present, then to support relay-assisted cooperative transmission,
both downlink and uplink time slots can be further divided into two
sub-slots, as shown in Fig.~\ref{fig:benchmarks}(ii). Clearly, this
four-time-slot TDD protocol is not efficient since resources will be
wasted when not every MS needs RS's assistance in both downlink and
uplink. In \cite{XuINFOCOM09}, the authors propose a transmission
protocol in FDD mode, where the total subcarriers are orthogonally
divided into data subcarrier pool and relay subcarrier pool.
However, in this FDD mode based transmission protocol, it is not
mentioned how to define a subcarrier as a data or relay subcarrier
and to determine the size of the two pools.

Inspired by the two-way relaying protocols studied in
\cite{PopovskiICC07} and \cite{KimIT08}, we propose a novel
three-time-slot TDD trasmission protocol as shown in
Fig.~\ref{fig:benchmarks}(iii), which can support the three
transmission modes, namely direct transmission, one- and two-way
relaying in the considered bidirectional cooperative cellular
networks. Specifically, in the first time slot, the BS transmits all
downlink signals while MSs and RSs listen. In the second time slot,
each MS transmits its uplink signals while the BS and RSs listen. In
the third time slot, RSs forward both downlink and uplink signals
received in the previous two time slots, whenever needed, while BS
and MSs listen. Thanks to the use of OFDMA,
the data streams for different MSs are transmitted on different
subcarriers in each time slot so that there is no multiple-access
interference.
%

%
The proposed transmission protocol can easily accommodate different
transmission modes in a unified fashion. For instance, if direct
transmission is preferred for a MS in both downlink and uplink,
then, bidirectional communication for this MS can be accomplished in
the first two time slots. If relay-assisted transmission is
preferred by a MS in both downlink and uplink, then a RS who
successfully decodes both the downlink and uplink messages can
combine the messages together using network coding and then send it
to both the BS and the MS in the third time slot.
More specifically, we list in Table~\ref{pairing} all the possible
combinations of transmission modes for each MS with bidirectional
traffic. As we can see, although both downlink and uplink traffic
can adopt one of the three transmission modes, and there are nine
combinations in total, but only five are feasible and marked as
``$\surd$". The other four are infeasible and marked as ``$\times$".
This is because two-way relaying can only take place when both
downlink and uplink transmissions need RS assistance. In other
words, the two-way relaying requires \emph{traffic pairing}. A
counterexample is the combination of ``two-way" in downlink and
``direct" in uplink, which obviously can never happen by definition.
A point worthwhile to mention is that such traffic pairing was
ignored in the previous work \cite{XuINFOCOM09}. From
Table~\ref{pairing} it is also seen that the combination of
``one-way" for uplink and ``one-way" for downlink is feasible. This
can be implemented by using two different RSs to forward downlink
and uplink transmissions, respectively. That is, both downlink and
uplink transmissions need RS assistance but do not necessarily use
network coding.

\begin{figure}[tbhp]
\begin{centering}
\includegraphics{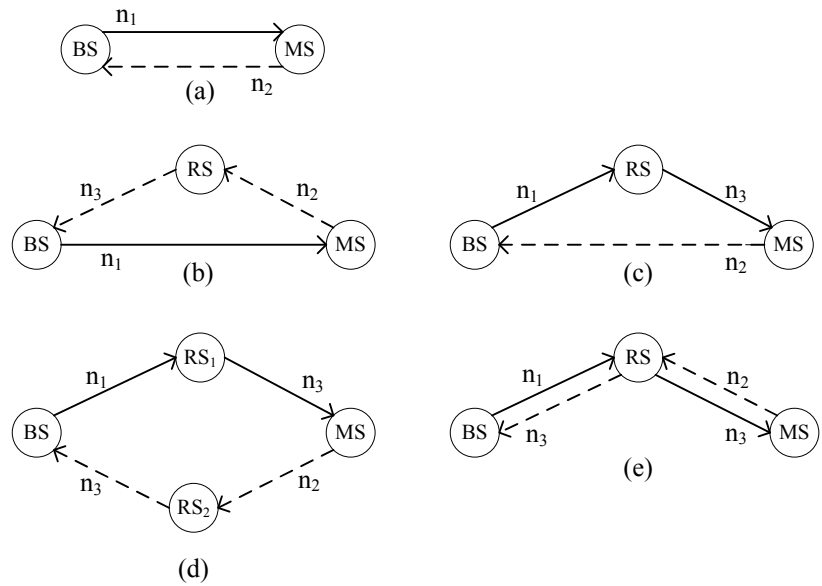}
\vspace{-0.1cm}
 \caption{Five feasible transmission modes. The solid lines represent downlink, while dashed lines represent uplink.}\label{transmission modes}
\end{centering}
\vspace{-0.3cm}
\end{figure}

Fig.~\ref{transmission modes} further illustrates all the five
feasible transmission modes, which will be considered throughout
this paper. In the figure, $n_t$ denotes the index of subcarrier
used for transmission in time slot $t$, for $t=1$,$2$ and $3$. By
introducing $n_t$, we gain the flexibility of adaptive subcarrier
assignment.  Note that in transmission mode $d$, both RSs occupy the
same subcarrier $n_3$ in the third time slot. This is feasible
because the back-propagated self-interference can be canceled in the
same way as in transmission mode $e$ for two-way relaying. Thus,
higher spectral efficiency can be obtained compared with the case
where the two RSs use different subcarriers.

\newtheorem{remark}{Remark}
\begin{remark}
In relay-assisted communications, the two received copies of the
same content at the destination, one from the source through the
direct link and the other from the relay through the cooperative
link, can be combined using maximum ratio combining (MRC). In this
paper, for simplicity, we assume that selection combining (SC) is
employed between the direct link and the cooperative link.
Therefore, every downlink and uplink traffic pair for each MS can
select one of the five transmission modes shown in
Fig.~\ref{transmission modes} according to the channel conditions.
\end{remark}

The proposed three-time-slot TDD transmission protocol can capture
the following gains:

\begin{itemize}
\item Cooperative diversity gain:
The relaying takes place in the third time slot only if needed and
each MS can select one or multiple RSs from all the available RSs in
the network.
%
%
%
\item Network coding gain:
For each MS, the uplink and downlink traffic always occurs in pair
so that we can enjoy the network coding gain as large as possible
through transmission mode selection as defined above.
\item Multiuser diversity gain: Subcarriers can be assigned adaptively to
different MSs in each time slot.
\end{itemize}

Before we propose in the next section an optimization framework that
simultaneously achieves the three kinds of gains, we need to make
the following assumptions in this paper.

First, it is assumed that full channel state information (CSI) of
the network is available at a central controller (which can be
embedded with the BS) and the transmission rate on each link can be
adapted based on it. Second, unlike the previous work
\cite{YuJSAC07, LiangVT09, XuINFOCOM09} where power allocation is
taken into account in the resource allocation, in this work we do
not pursue power allocation for simplicity. It is known that power
allocation can bring significant improvement in relay networks when
the source and relay nodes are subject to a total power constraint
\cite{Hammerstrom06}. However, as also demonstrated in \cite{Tao10,
Hammerstrom06, Ho08}, the gain brought by power adaptation is very
limited in OFDM-based relay networks if each transmitting node is
subject to an individual peak power constraint. In our considered
system model, all the BS, MSs and RSs are subject to their own
individual peak power constraints and, therefore, the transmit power
is assumed to be fixed and uniformly distributed among all
subcarriers for each of them.

Finally, we assume that the signal relaying is done per-subcarrier
basis. That is, the signal received on one subcarrier, say $i$, in
the first hop will be forwarded on subcarrier $i'$ in the next hop,
where the subcarrier index $i'$ may not be the same as $i$. This is
known as \textit{subcarrier-pairing} \cite{Tao10,Hammerstrom06} or
\textit{tone-permutation} \cite{Ho08}. Such subcarrier-pair based
relaying is optimal for amplified-and-forward (AF) protocol, where
the signals received by the same relay on different subcarriers are
processed individually, but suboptimal for decode-and-forward (DF)
protocol, where the information from one set of subcarriers in the
first hop can be decoded and re-encoded jointly and then transmitted
over a different set of subcarriers in the next hop. Nevertheless,
we still adopt the subcarrier-pair based relaying for simplicity. As
a result, the same number of subcarriers will be assigned in both
hops.

\section{Optimization Framework}

In this section, we present the optimization framework in details.
We first review the achievable downlink and uplink rate pair for
each feasible transmission mode.  Different relaying strategies
including AF and DF will be considered. Then we provide a rigorous
discussion of the problem formulation.

\subsection{Achievable Downlink and Uplink Rate Pairs}

Here we briefly discuss the rate expression for each of the five
transmission modes given in Fig.~\ref{transmission modes} to
facilitate the problem formulation in the next subsection.  We model
the wireless fading environment by large-scale path loss and
shadowing, along with small-scale frequency-selective Rayleigh
fading. OFDM is used at the physical layer and each subcarrier is
assumed to experience flat fading. We also assume that the channels
between different links experience independent fading. We further
assume that the network operates in slow fading environment, so that
channel estimation is perfect. The additive white Gaussian noises at
BS, RSs and MSs are assumed to be independent circular symmetric
complex Gaussian random variables. For brevity of notation,
subscripts $B$, $M$ and $R$ denote BS, MS and RS, respectively, $u$
and $d$ denote uplink and downlink, respectively.

\subsubsection{Transmission mode a}
In this mode, both downlink and uplink use direct transmission. The
achievable rate pair is easily obtained as
\begin{eqnarray}
R_d &=& \frac{1}{3}C(\gamma_{BM}),\\
R_u &=& \frac{1}{3}C(\gamma_{MB}),
\end{eqnarray}
where $C(x)={\rm log_2}(1+x)$, the pre-log factor $\frac{1}{3}$ is
due to the use of three time slots, and $\gamma_{ij}$ denotes the
signal-to-noise ratio (SNR) from the terminal $i$ to the terminal
$j$, for $i,j\in \{B, R, M\}$.

\subsubsection{Transmission mode b}

In this mode, the downlink traffic prefers direct transmission and
the uplink traffic needs RS assistance.
Currently, many relay strategies are proposed. Among them, the two
popular and practical ones are known as AF and DF. We thus focus on
AF and DF throughout this paper. Then, we can write the achievable
rate pair as
\begin{eqnarray}
R_d  &=&  \frac{1}{3}C(\gamma_{BM}),\\
R_u  &=&  \begin{cases}
\frac{1}{3}C\left(\frac{\gamma_{MR}\gamma_{RB}}{1+\gamma_{MR}+\gamma_{RB}}\right),
& {\rm for ~AF} \\
\frac{1}{3}{\rm min}\left\{C(\gamma_{MR}), C(\gamma_{RB})\right\}.
\label{eqn:DF} & {\rm for ~DF}
\end{cases}
\end{eqnarray}
%

\subsubsection{Transmission mode c}
In this case, the uplink traffic prefers direct transmission and the
downlink traffic requires RS assistance. The achievable rate pair
can be similarly rewritten as
\begin{eqnarray}
R_d & = & \begin{cases}
\frac{1}{3}C\big(\frac{\gamma_{BR}\gamma_{RM}}{1+\gamma_{BR}+\gamma_{RM}}\big),
&{\rm for ~AF} \\
\frac{1}{3}{\rm min}\{C(\gamma_{BR}), C(\gamma_{RM})\}, &{\rm
for~DF}
\end{cases}\\
R_u  &=&  \frac{1}{3}C(\gamma_{MB}).
\end{eqnarray}
\subsubsection{Transmission mode d}
In this case, both downlink and uplink traffic needs RS assistance
but via two different RSs over the same subcarrier. The downlink and
uplink signals from two RSs will both arrive at BS and MS, resulting
in inter-link interference. It can be easily verified that the
interference can be completely canceled since they are the
back-propagated self-interference from BS or MS's \textit{priori}
transmission. A special note is that in the case of AF, each
destination also receives the amplified noises from both RSs which
cannot be canceled. Thus, the achievable rate pair can be obtained
as, whose derivation is simple and ignored.
\begin{eqnarray}
R_d & = &\begin{cases}
\frac{1}{3}C\left(\frac{\gamma_{BR_1}\gamma_{R_1M}(1+\gamma_{MR_2})}{\gamma_{R_1M}(1+\gamma_{MR_2})
+ \gamma_{R_2M}(1+\gamma_{BR_1}) +
(1+\gamma_{MR_2})(1+\gamma_{BR_1})}\right),
&{\rm for ~AF}\\
\frac{1}{3}{\rm min}\{C(\gamma_{BR_1}), C(\gamma_{R_1M})\}, &{\rm
for
~DF}\end{cases}\\
R_u & = &\begin{cases}
\frac{1}{3}C\left(\frac{\gamma_{MR_2}\gamma_{R_2B}(1+\gamma_{BR_1})}{\gamma_{R_1B}(1+\gamma_{MR_2})
+ \gamma_{R_2B}(1+\gamma_{BR_1}) +
(1+\gamma_{MR_2})(1+\gamma_{BR_1})}\right), &{\rm for ~AF}\\
\frac{1}{3}{\rm min}\{C(\gamma_{MR_2}), C(\gamma_{R_2B})\}. &{\rm
for ~DF}\end{cases}
\end{eqnarray}

\subsubsection{Transmission mode e}
This is the 3-step two-way relaying, where BS transmits its signals
to RS in the first time slot, MS transmits its signals to
 RS in the second time slot, RS then mixes the received
signals and broadcasts it to both BS and MS in the third time slot.
Depending on if AF or DF is used, we present the achievable rate
pairs separately in what follows.

%

\paragraph{AF two-way relaying}
The achievable rate pair for 2-step AF two-way relaying is studied
in \cite{RankovJSAC07, PopovskiICC07}. The extension to the 3-step
protocol is simple. The results are:
\begin{eqnarray}
R_d &=&
\frac{1}{3}C\left(\frac{\alpha^2\gamma_{BR}|h_{RM}|^2}{1+(\alpha^2+\beta^2)|h_{RM}|^2}\right),\label{eqn:PNC-AFd}
\\
R_u &=&
\frac{1}{3}C\left(\frac{\beta^2\gamma_{MR}|h_{RB}|^2}{1+(\alpha^2+\beta^2)|h_{RB}|^2}\right),\label{eqn:PNC-AFu}
\end{eqnarray}
where
\begin{equation}\label{eqn:scale}
{\alpha=\sqrt{\frac{\xi
P_R}{1+\gamma_{BR}}}},~~~~~~{\beta=\sqrt{\frac{(1-\xi)P_R}{1+\gamma_{MR}}}}.
\end{equation}
Here, $\xi\in [0,1]$ is a power allocation coefficient that
determines the weights of the signals from BS and MS in the combined
signals, $P_R$ is the transmit power constraint at RS, and $h_{i,j}$
is the channel gain from terminal $i$ to terminal $j$.

\paragraph{DF two-way relaying}
After the RS decodes the messages from the BS and MS, it can combine
the messages using either bitwise XOR or symbol-based superposition
(SUP).
For bitwise XOR, the rate pair is given by \cite{PopovskiICC07}:
\begin{eqnarray}
R_d &=& \frac{1}{3}{\rm min}\{C(\gamma_{BR}), C(\gamma_{RB}), C(\gamma_{RM})\}, \label{eqn:XOR} \\
R_u &=& \frac{1}{3}{\rm min}\{C(\gamma_{MR}), C(\gamma_{RB}),
C(\gamma_{RM})\}, \label{eqn:XOR'}
\end{eqnarray}
where the first term in (\ref{eqn:XOR}) or (\ref{eqn:XOR'})
represents the maximum rate at which RS can reliably decode the
signals from BS or MS, while the minimum of the second and third
terms in both (\ref{eqn:XOR}) and (\ref{eqn:XOR'}) represents the
maximum rate at which both BS and MS can reliably decode the signals
from RS during the broadcast phase.

\paragraph{SUP-based DF two-way relaying}

If SUP is applied, the rate pair is easily obtained as
\cite{RankovJSAC07, OechteringTC08}
%
\begin{eqnarray}
R_d &=& \frac{1}{3}{\rm min}\{C(\gamma_{BR}), C(\theta\gamma_{RM})\},\label{eqn:SUP} \\
R_u &=& \frac{1}{3}{\rm min}\{C(\gamma_{MR}),
C((1-\theta)\gamma_{RB})\},\label{eqn:SUP'}
\end{eqnarray}
where $\theta\in [0,1]$ is a power allocation coefficient.


\begin{remark}
In all the aforementioned rate pair expressions, the SNR
$\gamma_{ij}$ may not be the same as $\gamma_{ji}$. This is not only
because the transmit power on terminal $i$ and $j$ may be different,
but more importantly, the two links $i\rightarrow j$ and
$j\rightarrow i$ can be assigned two different physical subcarriers.

\end{remark}

\subsection{Problem Formulation}
Let $\mathcal {K}=\{1,2,...,K\}$ be the set of MSs, $\mathcal
{M}=\{1,2,...,M\}$ the set of RSs and $\mathcal {N}=\{1,2,...,N\}$
the set of subcarriers.
%
%
%
Among the five transmission modes shown in Fig.~\ref{transmission
modes}, the first mode (direct transmission) involves subcarrier
assignment in the first and second time slots only. For the rest
four modes (cooperative transmission), one needs to assign the
subcarriers in all the three time slots as well as selecting the
proper RS(s). On the other hand, as discussed earlier, the downlink
and uplink data streams are paired and can take the five possible
transmission modes. In view of these facts,
%
%
we introduce the following five sets of binary variables for
transmission mode selection:
\begin{itemize}
\item[-] $\rho_{k,a}^{n_1,n_2}$ indicates whether subcarrier pair
$(n_1,n_2)$ in the first two time slots is assigned for MS $k$ for
direct transmission of downlink and uplink traffic using
transmission mode $a$, where $n_t$ denotes the subcarrier in time
slot $t$.
\item[-] $\rho_{k,r,b}^{n_1,n_2,n_3}$ indicates whether MS $k$ is assigned
RS $r$ for uplink traffic on subcarriers $n_2$ and $n_3$ and direct
transmission of downlink traffic on subcarrier $n_1$, using
transmission mode $b$.
\item[-] $\rho_{k,r,c}^{n_1,n_2,n_3}$ indicates whether MS $k$ is
assigned RS $r$ for downlink traffic on subcarriers $n_1$ and $n_3$
and direct transmission of uplink traffic on subcarrier $n_2$, using
transmission mode $c$.
\item[-] $\rho_{k,r,r',d}^{n_1,n_2,n_3}$ indicates whether MS
$k$ is assigned RS $r$ for downlink traffic on subcarriers $n_1$ and
$n_3$ and RS $r'$ for uplink traffic on subcarriers $n_2$ and $n_3$,
using transmission mode $d$.
\item[-] $\rho_{k,r,e}^{n_1,n_2,n_3}$ indicates
whether MS $k$ is assigned RS $r$ for downlink traffic on
subcarriers $n_1$ and $n_3$ and uplink traffic on subcarriers $n_2$
and $n_3$, using transmission mode $e$.
\end{itemize}

In this paper we assume that each subcarrier in each time slot can
only be assigned to one MS for one traffic session in order to avoid
interference. Moreover, the traffic session on each subcarrier can
only operate in one of the five transmission modes. Therefore,
%
%
%
these binary variables must satisfy the constraints:
\begin{eqnarray}\label{eqn:C2}
\sum_{\begin{subarray}{c}k\in\mathcal
{K}\\n_2\in\mathcal {N}\end{subarray}}\rho_{k,a}^{n_1,n_2}+\sum_{\begin{subarray}{c} k\in\mathcal {K},r\in\mathcal {M}\\
n_2\in\mathcal {N},n_3\in\mathcal
{N}\end{subarray}}\Big(\rho_{k,r,b}^{n_1,n_2,n_3}+\rho_{k,r,c}^{n_1,n_2,n_3}+\sum_{\begin{subarray}{c}r'\in\mathcal
{M}\\r'\neq r\end{subarray}}\rho_{k,r,r',d}^{n_1,n_2,n_3}
+\rho_{k,r,e}^{n_1,n_2,n_3}\Big)\leq1, \forall n_1\in\mathcal {N},
\end{eqnarray}
\begin{eqnarray}\label{eqn:C3}
\sum_{\begin{subarray}{c}k\in\mathcal
{K}\\n_1\in\mathcal {N}\end{subarray}}\rho_{k,a}^{n_1,n_2}+\sum_{\begin{subarray}{c} k\in\mathcal {K},r\in\mathcal {M}\\
n_1\in\mathcal {N},n_3\in\mathcal
{N}\end{subarray}}\Big(\rho_{k,r,b}^{n_1,n_2,n_3}+\rho_{k,r,c}^{n_1,n_2,n_3}
+\sum_{\begin{subarray}{c}r'\in\mathcal {M}\\r'\neq
r\end{subarray}}\rho_{k,r,r',d}^{n_1,n_2,n_3}
+\rho_{k,r,e}^{n_1,n_2,n_3}\Big)\leq1, \forall n_2\in\mathcal {N},
\end{eqnarray}
\begin{eqnarray}\label{eqn:C4}
\sum_{\begin{subarray}{c} k\in\mathcal {K},r\in\mathcal {M}\\
n_1\in\mathcal {N},n_2\in\mathcal
{N}\end{subarray}}\Big(\rho_{k,r,b}^{n_1,n_2,n_3}+\rho_{k,r,c}^{n_1,n_2,n_3}
+\sum_{\begin{subarray}{c}r'\in\mathcal {M}\\r'\neq
r\end{subarray}}\rho_{k,r,r',d}^{n_1,n_2,n_3}
+\rho_{k,r,e}^{n_1,n_2,n_3}\Big)\leq1, \forall n_3\in\mathcal {N}.
\end{eqnarray}

After introducing these variables, we can now characterize the
achievable downlink-uplink sum rate of each MS $k$ over all the
possible transmission modes. This is given by
\begin{eqnarray}\label{eqn:C1}
R_k^{\rm sum} &=& \sum_{n_1\in\mathcal {N}}\sum_{n_2\in\mathcal
{N}}{\Big(R_{k,d}^{n_1}+R_{k,u}^{n_2}\Big)\rho_{k,a}^{n_1,n_2}} \nonumber \\
&&  + \sum_{n_1\in\mathcal {N}}\sum_{n_2\in\mathcal
{N}}\sum_{n_3\in\mathcal {N}}\sum_{r\in\mathcal
{M}}{\Big(R_{k,d}^{n_1}+R_{k,u}^{r,n_2,n_3}\Big)\rho_{k,r,b}^{n_1,n_2,n_3}} \nonumber \\
&& + \sum_{n_1\in\mathcal {N}}\sum_{n_2\in\mathcal
{N}}\sum_{n_3\in\mathcal {N}}\sum_{r\in\mathcal
{M}}{\Big(R_{k,d}^{r,n_1,n_3}+R_{k,u}^{n_2}}\Big)\rho_{k,r,c}^{n_1,n_2,n_3} \nonumber \\
&& + \sum_{n_1\in\mathcal {N}}\sum_{n_2\in\mathcal
{N}}\sum_{n_3\in\mathcal {N}}\sum_{r\in\mathcal
{M}}\sum_{\begin{subarray}{c}r'\in\mathcal {M}\\r'\neq
r\end{subarray}}{\Big(R_{k,d}^{r,n_1,n_3}+R_{k,u}^{r',n_2,n_3}\Big)\rho_{k,r,r',d}^{n_1,n_2,n_3}} \nonumber \\
&& + \sum_{n_1\in\mathcal {N}}\sum_{n_2\in\mathcal
{N}}\sum_{n_3\in\mathcal {N}}\sum_{r\in\mathcal
{M}}{\Big(R_{k,d}^{r,n_1,n_3}+R_{k,u}^{r,n_2,n_3}}\Big)\rho_{k,r,e}^{n_1,n_2,n_3}.~~~~
\end{eqnarray}
The five summation terms in (\ref{eqn:C1}) represent the
downlink-uplink sum rate achieved over the five transmission modes,
respectively. The detailed rate expressions can be found in the
previous subsection. In addition, the number of non-zero elements in
each summation term represents the number of downlink-uplink traffic
sessions that take on the same transmission mode but over different
sets of subcarriers for this MS.
From (\ref{eqn:C1}), one can also find that each MS can
simultaneously operate in all the five transmission modes.

Our objective is to maximize the  system total throughput by not
only allocating subcarriers optimally but also finding the best RSs
and best transmission modes for each MS. This is formulated as
follows (P1):
\begin{eqnarray}\label{eqn:obj}
&& {{\rm max}~R_{tot}=\sum_{k\in\mathcal {K}}{R_{k}^{\rm sum}} } \nonumber \\
&& \textit{s.t.} ~~ (\ref{eqn:C2}),(\ref{eqn:C3}),(\ref{eqn:C4}).
\end{eqnarray}
\begin{remark}
For bidirectional cellular networks, there is no single figure of
merit to measure the overall system performance. For simplicity, we
choose the downlink-uplink sum rate as our objective function.
Notice that, if the asymmetric traffic in the downlink and uplink is
considered, we can easily change to weighted sum rate, where the
weighting parameters can be adjusted to accommodate the asymmetry
requirement. In addition, we can also easily modify the objective
function to weighted sum of MS's rate if user fairness is concerned.
\end{remark}

\section{A Graph-Based Approach}
Problem P1 is a combinatorial optimization problem and looks
formidable at a first glance as it involves too many binary
variables.
%
%
Conventional convex optimization techniques such as Lagrange dual
decomposition as used in \cite{YuJSAC07,LiangVT09} cannot solve it
efficiently. Other optimization approaches, such as cutting plane
and branch-and-bound algorithms \cite{Papadimitriou}, are also not
viable due to the prohibitively large complexity. In this section,
we propose a graph-based metaheuristic approach to solve the
optimization problem. We first establish and prove the equivalence
between the original optimization problem and a MWCP in graph
theory. We then propose an ACO algorithm that runs in polynomial
time.

\subsection{Graph Model}
Let us rewrite the system total sum rate as:
\begin{eqnarray}\label{eqn:Ctot}
R_{tot} &=& \sum_{n_1\in\mathcal {N}}\sum_{n_2\in\mathcal
{N}}\sum_{k\in \mathcal {K}}{\Big(R_{k,d}^{n_1}+R_{k,u}^{n_2}\Big)\rho_{k,a}^{n_1,n_2}} \nonumber \\
&&  + \sum_{n_1\in\mathcal {N}}\sum_{n_2\in\mathcal
{N}}\sum_{n_3\in\mathcal {N}}\sum_{k\in \mathcal
{K}}\sum_{r\in\mathcal
{M}}\Bigg\{{\Big(R_{k,d}^{n_1}+R_{k,u}^{r,n_2,n_3}\Big)\rho_{k,r,b}^{n_1,n_2,n_3}} \nonumber \\
&&
+{\Big(R_{k,d}^{r,n_1,n_3}+R_{k,u}^{n_2}}\Big)\rho_{k,r,c}^{n_1,n_2,n_3}
 + \sum_{\begin{subarray}{c}r'\in\mathcal {M}\\r'\neq
r\end{subarray}}{\Big(R_{k,d}^{r,n_1,n_3}+R_{k,u}^{r',n_2,n_3}\Big)\rho_{k,r,r',d}^{n_1,n_2,n_3}} \nonumber \\
&& +
{\Big(R_{k,d}^{r,n_1,n_3}+R_{k,u}^{r,n_2,n_3}}\Big)\rho_{k,r,e}^{n_1,n_2,n_3}\Bigg\}.
\end{eqnarray}
Observing the first summation term of (\ref{eqn:Ctot}), it is easy
to find that there is at most one non-zero element for a given
subcarrier set $(n_1,n_2)$ due to the constraints (\ref{eqn:C2}) to
(\ref{eqn:C4}). This implies that among the $K$ MSs, at most one MS
can occupy the subcarrier tuple $(n_1,n_2)$ for direct transmission.
Similarly, observing the second summation term of (\ref{eqn:Ctot}),
we find that there is also at most one non-zero element for a given
subcarrier set $(n_1,n_2,n_3)$. This implies that at most one MS can
occupy the subcarrier tuple $(n_1,n_2,n_3)$ for transmission using
only one of the four relay-assisted transmission modes.

Based on the above observation, we can define
\begin{eqnarray}\label{eqn:k*}
\mathcal {R}(n_1,n_2) = \max_{k\in\mathcal
{K}}(R_{k,d}^{n_1}+R_{k,u}^{n_2}),
\end{eqnarray}
for each possible subcarrier pair $(n_1,n_2)$, and
\begin{eqnarray}\label{eqn:mpr}
 \mathcal {R}(n_1,n_2,n_3) =
\max_{k\in\mathcal {K}}\max_{r\in\mathcal
{M}}\Big\{(R_{k,d}^{n_1}+R_{k,u}^{r,n_2,n_3}),(R_{k,d}^{r,n_1,n_3}+R_{k,u}^{n_2}), \nonumber \\
 \max_{\begin{subarray}{c}r'\in\mathcal {M}\\r'\neq
r\end{subarray}}(R_{k,d}^{r,n_1,n_3}+R_{k,u}^{r',n_2,n_3}),(R_{k,d}^{r,n_1,n_3}+R_{k,u}^{r,n_2,n_3})\Big\},
\end{eqnarray}
for each possible subcarrier tuple $(n_1,n_2,n_3)$. Then, to
maximize the system total throughput, $R_{tot}$ can be represented
without loss of optimality as
\begin{equation}\label{eqn:tot'}
R_{tot}'=\sum_{n_1\in\mathcal {N}}\sum_{n_2\in\mathcal {N}}\mathcal
{R}(n_1,n_2)\rho_{k^*,a}^{n_1,n_2}+\sum_{n_1\in\mathcal
{N}}\sum_{n_2\in\mathcal {N}}\sum_{n_3\in\mathcal {N}}\mathcal
{R}(n_1,n_2,n_3)\rho_{k^*,p^*,\Omega^*}^{n_1,n_2,n_3},
\end{equation}
where $k^*$ in the first summation term represents the MS index that
takes the maximum in (\ref{eqn:k*}), and $\{k^*, p^*, \Omega^*\}$
represent the MS index, transmission mode index and RS index,
respectively that takes the maximum in (\ref{eqn:mpr}). Note that
$\Omega=\{r,r'\}$ if $p=$ mode $d$ and $\Omega=r$ if $p=$ mode $b$,
$c$, and $e$. Accordingly, the constraints (\ref{eqn:C2}) to
(\ref{eqn:C4}) can be rewritten as
\begin{eqnarray}
&&\sum_{n_2\in\mathcal
{N}}\rho_{k^*,a}^{n_1,n_2}+\sum_{n_2\in\mathcal {N}}
\sum_{n_3\in\mathcal {N}}\rho_{k^*,p^*,\Omega^*}^{n_1,n_2,n_3}\leq1,
~~~\forall n_1\in\mathcal {N},\label{eqn:c1'} \\
&&\sum_{n_1\in\mathcal
{N}}\rho_{k^*,a}^{n_1,n_2}+\sum_{n_1\in\mathcal
{N}}\sum_{n_3\in\mathcal
{N}}\rho_{k^*,p^*,\Omega^*}^{n_1,n_2,n_3}\leq1,
~~~\forall n_2\in\mathcal {N},\label{eqn:c2'} \\
&& \sum_{n_1\in\mathcal {N}}\sum_{n_2\in\mathcal
{N}}\rho_{k^*,p^*,\Omega^*}^{n_1,n_2,n_3}\leq1, ~~~\forall
n_3\in\mathcal {N}.\label{eqn:c3'}
\end{eqnarray}
Consequently, we can transform the original problem P1 to the
following problem (P2):
\begin{eqnarray}\label{eqb:obj'}
&& {{\rm max}~R_{tot}' } \nonumber \\
&& \textit{s.t.}~~(\ref{eqn:c1'}),(\ref{eqn:c2'}),(\ref{eqn:c3'})
\end{eqnarray}

The simplified problem P2 involves $N^2+N^3$ binary variables only.
It can be readily solved by the branch-and-bound algorithm, the
well-known method for finding the optimal solution to combinatorial
problems \cite{Papadimitriou}. However, its \textit{potential}
computational complexity still grows exponentially with $N^2+N^3$.
This motivates us to seek a graphic approach for solving P2 in
polynomial time.

\begin{figure}[tbhp]
\begin{centering}
\includegraphics[scale=.8]{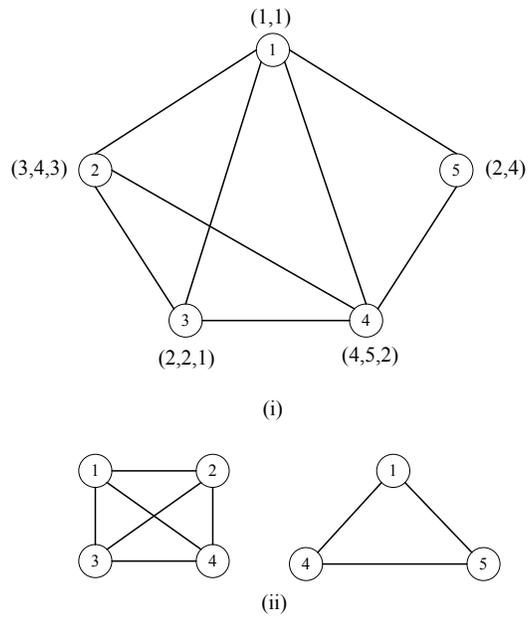}
\vspace{-0.1cm}
 \caption{Graph and clique example: (i) the graph; (ii)
two cliques.)}\label{clique}
\end{centering}
\vspace{-0.3cm}
\end{figure}

\newtheorem{theorem}{Theorem}
\begin{theorem}
Problem P2 is equivalent to a maximum weighted clique problem
(MWCP).
\end{theorem}
\begin{proof}
Let $\mathcal {G}=(\mathcal {V},\mathcal {E},\mathcal {W})$ be an
arbitrary undirected and weighted graph, where $\mathcal {V}$ is a
set of vertices, $\mathcal {E}\subseteq \mathcal {V}\times \mathcal
{V}$ is a set of edges, $\mathcal {W}$ is the weighting function
such that $\mathcal {W}: \mathcal {V}\rightarrow \mathbb{R}_+$. A
\textit{clique} is a set of vertices $\mathcal {C}\subseteq \mathcal
{V}$ such that every pair of distinct vertices of $\mathcal {C}$ is
connected with an edge. Based on the above discussion, we define two
type of vertices for the given problem. For type one, a
\textit{vertex} is a subcarrier pair $(n_1,n_2)$ in the first two
time slots associated with transmission mode $a$.
For type two, a vertex is a subcarrier tuple $(n_1, n_2, n_3)$
associated with the four relay-assisted transmission modes. Due to
the three-time-slot TDD mode, the set of subcarriers $\mathcal {N}$
is shared in each time slot. Consequently, the total number of
distinct vertices are $N^2+N^3$, i.e., $|\mathcal {V}|=N^2+N^3$,
where $|\cdot|$ is cardinality of a set. We define two vertices
\textit{intersect} if they have no common element in each time slot,
and \textit{disjoint} if they have at least one common element in
one time slot. For any pair of vertices that intersect, we connect
them by an edge. An example for the graph construction and clique is
shown in Fig.~\ref{clique}.

When the graph is constructed, each vertex is given a weight, which
is defined as the maximum achievable rate over the given subcarrier
tuple. Specifically,
\begin{eqnarray}
\mathcal{W}_{(n_1,n_2)}&=&\mathcal {R}(n_1,n_2), \\
\mathcal{W}_{(n_1,n_2,n_3)}&=&\mathcal {R}(n_1,n_2,n_3).
\end{eqnarray}
The above weighting process is to find the best MS for a given
vertex $(n_1,n_2)$ and the best combination of MSs, RSs and
transmission modes for a given vertex $(n_1,n_2,n_3)$, for
maximizing the achievable rate. Note that for each of the two type
of vertices $(n_1,n_2)$ and $(n_1,n_2,n_3)$, the complexity of
weighting process is $\mathcal {O}(K)$ and $\mathcal
{O}\left(3KM+KM(M-1)\right)$, respectively. Therefore, the total
complexity of weighting process is $\mathcal{O}(N^2K + 3N^3KM +
N^3KM(M-1))$.

Having defined the graph we now turn to the optimization problem
defined in P2. According to our construction methods of vertices and
edges, we find that the mutually adjacent vertices can be selected
simultaneously without violating the exclusive subcarrier assignment
in each time slot defined in (\ref{eqn:c1'}) to (\ref{eqn:c3'}). The
weighting process is done over all vertices in $\mathcal {V}$ so as
to match them to the corresponding optimal MSs or combinations of
MSs, RSs and transmission modes, maximizing their achievable rate.
Therefore, jointly optimizing the subcarrier assignment,
transmission mode selection and relay selection for system total
throughput maximization is to find a subset $\mathcal {C}$ of
pairwise adjacent vertices in the graph having the largest total
weight, i.e., the so called MWCP (P3):
\begin{equation}
{\max_{\mathcal {C} \subseteq \mathcal {V} } \mathcal {W}(\mathcal
{C})=\sum_{v\in\mathcal {C}}{\mathcal {W}_v}}.
\end{equation}
Therefore, P2 is equivalent to P3. The theorem is proved.
\end{proof}

\subsection{ACO for MWCP}
 Like the maximum clique problem (MCP) (finding a clique having
the largest cardinality), the MWCP is a classical
 combinatorial optimization problem and is NP-complete. In this section, we introduce
 an ACO algorithm  based metaheuristic to solve the MWCP.

 The ACO metaheuristic is a bio-inspired approach that has been used to
 solve different hard combinatorial optimization problems. The main
 idea of ACO is to model the problem as the search for a minimum
 cost path in a graph. Artificial ants walk through this graph,
 looking for good paths. Each ant has a rather simple behavior so
 that it will likely find rather poor quality paths on its
 own. Better paths are found as the emergent result of the global
 cooperation among ants in the colony. This cooperation is performed
 in an indirect way through pheromone laying \cite{Dorigo,Solnon1}.

 In \cite{Solnon1}, authors propose an ACO algorithm for solving the
 MCP. It requires only a trivial modification that transforms the cardinality function
 to the weight function for MWCP. The modification is commonly used
 for MCP and MWCP in the field of graph theory \cite{Balas1,Balas2}.
 The ACO algorithm for MWCP consists of three steps: pheromone trail
 initialization, construction of cliques by ants and updating
 pheromone trails. The details are sketched in Appendix \ref{app:aco}.

%

The total time complexity of ACO for WMCP is linear in $|\mathcal
{V}|$ and in the order of $\mathcal {O}\left({\rm
ite}\cdot(|\mathcal {V}|+\delta)\right)$ \cite{Solnon1}, where ${\rm
ite}$ is the maximum number of iterations obtained empirically and
$\delta$ is a constant (related to the number of ants, the size of
the maximum weighted clique and the maximum vertex degree in
$\mathcal {G}$). By combining the weighting complexity as mentioned
in the proof of Theorem 1, the overall computational complexity of
the proposed algorithm is thus given by $\mathcal{O}\big(N^2K +
3N^3KM + N^3KM(M-1)+{\rm ite}\cdot(N^2 + N^3 +\delta)\big)$, which
is polynomial in the system parameters $N$ (number of subcarriers),
$K$ (number of MSs) and $M$ (number of RSs).

\begin{remark}
The key of the proposed algorithm is the mapping from the original
problem P1, to the simplified P2 and then to the MWCP P3. Once the
mapping process is done, there exist many graphical methods for
solving the MWCP, such as simple greedy heuristic or complicated
reactive Tabu search. We choose the ACO algorithm because of its
ability to strike a good balance between performance and
computational complexity \cite{Solnon1}. Comparison of graphical
methods is out of the scope of this paper.
\end{remark}

\section{Simulation Results}
In this section, we evaluate the performance of the proposed
transmission protocol together with the ACO based resource
allocation algorithm using simulation.

\begin{figure}[tbhp]
\begin{centering}
\includegraphics[scale=.8]{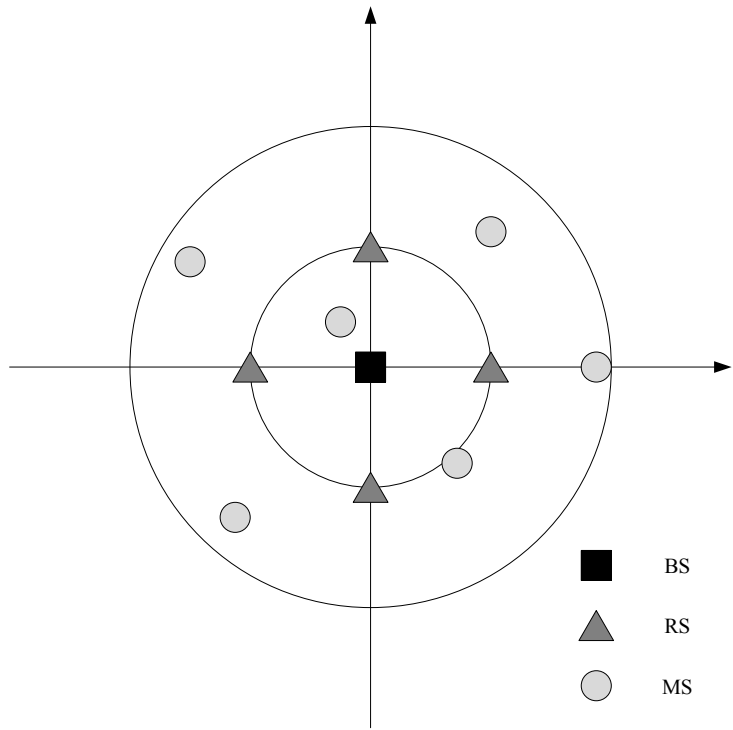}
\vspace{-0.1cm}
 \caption{Two-dimensional plan of nodes location.}\label{fig:setup}
\end{centering}
\vspace{-0.3cm}
\end{figure}

We consider a cell with $2$ km radius, where MSs are uniformly
distributed in the cell and RSs are uniformly distributed on a
circle centered at the BS and with radius of 1 km. The corresponding
two-dimensional plane is shown in Fig.~\ref{fig:setup}. The central
frequency is around 5 GHz. The statistical path loss model and
shadowing are referred to \cite{Erceg99}, where we set the path loss
exponent to be $4$ and the standard deviation of log-normal
shadowing is $5.8$ dB. The small-scale fading is modeled by Rayleigh
fading process, where the power delay profile is exponentially
decaying with maximum delay spread of 5 $\mu s$ and maximum Doppler
spread of 30 Hz. A total of $2000$ different channel realizations
were used. For each channel realization, the locations of the MSs
are random but uniform distributed.
For illustration purpose, the total number of subcarriers is $N=16$.
All RSs have the same maximum power constraints,  so do all MSs.
We consider that the maximum power constraints in dB at BS, RS and
MS satisfy $P_B=P_R+3{\rm dB}=P_M+5{\rm dB}$. For simplicity, the
power allocation coefficients in AF and SUP-based DF two-way
relaying are $\xi=0.5$ and $\theta=0.5$, respectively. The graph
settings in the ACO algorithm are listed in Table~\ref{tab:graph},
which have been proved to be efficient in \cite{Solnon1}.

\subsection{A toy example}

\begin{figure}[tbhp]
\begin{centering}
\includegraphics[scale=.8]{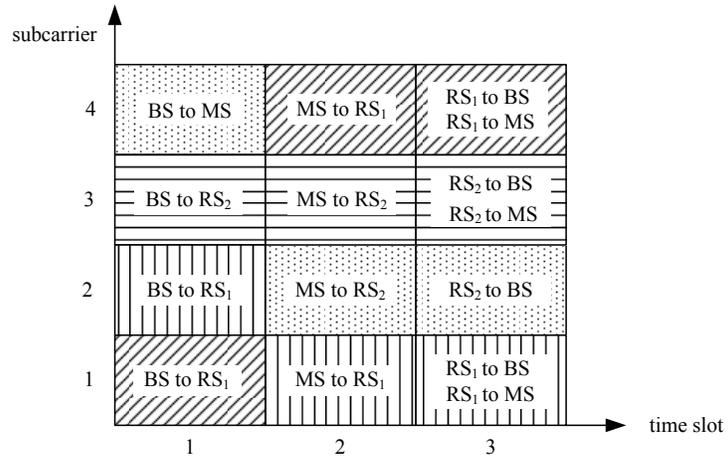}
\vspace{-0.1cm}
 \caption{Example for the proposed algorithm. Each pattern represents one downlink/uplink traffic session pair.}\label{fig:example}
\end{centering}
\vspace{-0.3cm}
\end{figure}

To vividly describe the proposed algorithm, we first consider a toy
example with one MS, two RSs, and four subcarriers. For notation
convenience, we transform the cell into the equivalent
two-dimensional plan with radius of 10, where the BS is fixed at
$(0,0)$, the MS is fixed at $(10,0)$, and the two RSs are fixed at
$(4,3)$ and $(4,-3)$, respectively. This is without loss of
generality and commonly used in \cite{YuJSAC07} and
\cite{LiangVT09}. The per-subcarrier power constraint of BS is 10
dB, XOR-based DF two-way relaying is used (Note that the RS's
operation is also DF in the case of one-way relaying). For a given
channel realization, applying our proposed algorithm, we obtain a
maximum weighted clique consisting of four vertices
$\{(1,4,4),(2,1,1),(3,3,3),$ $(4,2,2)\}$. The corresponding weights
are $\{0.1361,0.1341,0.0780,0.0407\}$ (bits/s/Hz), and the
corresponding transmission modes are $\{e,e,e,b\}$, respectively.
The subcarrier assignment, relay selection and transmission mode
selection are conveniently represented using the time-frequency grid
shown in Fig.~\ref{fig:example}.

\subsection{Comparison with benchmark schemes}

We now evaluate the performance of our proposed three-time-slot TDD
transmission protocol in comparison with the two benchmark schemes
(denoted as BM1 and BM2, respectively) shown in
Fig.~\ref{fig:benchmarks}.
Unlike the proposed protocol, the uplink and downlink optimization
for the benchmark schemes is decoupled as there is no correlation
between them.
For BM1, we use a greedy policy that assigns every subcarrier to the
MS with the best channel condition for both downlink and uplink,
respectively, which is optimal for throughput optimization.
For BM2, since the optimal solution is difficult to obtain for
either downlink or uplink throughput optimization, we similarly map
them into a MWCP and solve by ACO. Specifically, for downlink
traffic, we use subcarrier tuples $(n_1)$ and $(n_1,n_2)$ represent
the vertices for direct transmission and one-way relaying,
respectively. Weighting is done across MSs or all different
combinations of RSs and MSs, and disjoint vertices is selected by
ACO algorithm. Similar process is done for uplink traffic.

\begin{figure}[tbhp]
\begin{centering}
\includegraphics[scale=.8]{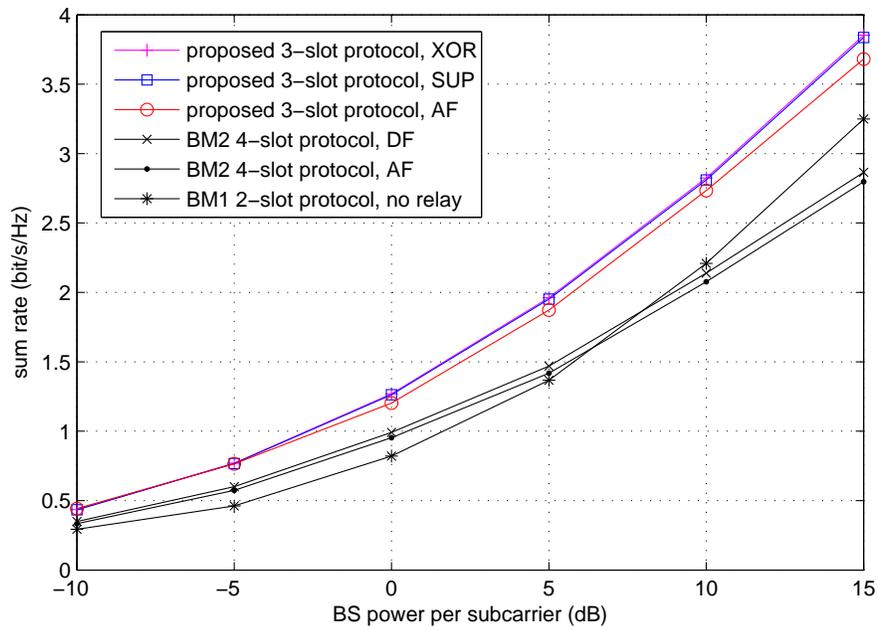}
\vspace{-0.1cm}
 \caption{Performance comparison of the proposed transmission protocol and two benchmarks.}\label{fig:throughput1}
\end{centering}
\vspace{-0.3cm}
\end{figure}

Fig.~\ref{fig:throughput1} shows the results when there are $K=4$
MSs, $M=10$ RSs and $N=16$ subcarriers.
From the figure, we first observe that BM2 (with conventional relay)
only slightly outperforms BM1 (no relay) when SNR is below around
$8$ dB and is inferior to BM1 when SNR is higher. This observation
suggests that conventional relaying is not always helpful in
cellular networks.
%
%
It is also observed that our proposed transmission protocol
outperforms both BM1 and BM2 substantially over a wide range of SNR.
In particular, compared with BM1, about 20\% and 30\% throughput
improvements are achieved when AF and DF are used in our protocol,
respectively. This clearly demonstrates the superiority of our
proposed three-time-slot transmission protocol by making the best
use of cooperative diversity and network coding gain. In addition,
among the three two-way relaying strategies, we find that the two DF
strategies perform close to each other while the AF strategy is
slightly worse.
%
%
%

\subsection{Comparison with different adaptation schemes}

\begin{figure}[tbhp]
\begin{centering}
\includegraphics[scale=.8]{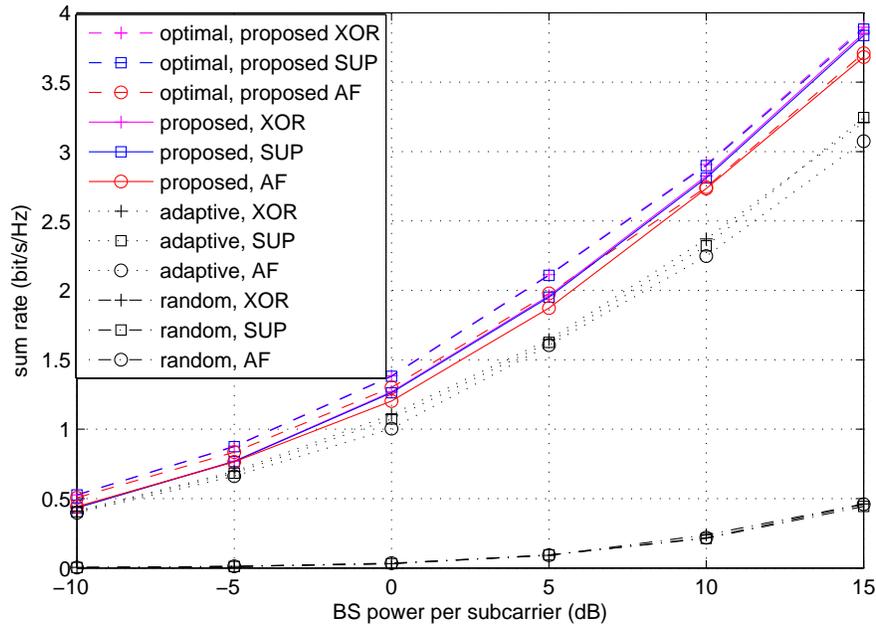}
\vspace{-0.1cm}
 \caption{Performance comparison of the proposed algorithm and two suboptimal resource allocation schemes.}\label{fig:throughput2}
\end{centering}
\vspace{-0.3cm}
\end{figure}

In this subsection, we demonstrate the efficiency of the proposed
ACO based adaptive resource allocation algorithm over two suboptimal
resource allocation schemes in Fig.~\ref{fig:throughput2}. The
optimal solution obtained by the branch-and-bound
algorithm\footnote{The branch-and-bound method is implemented by
``bintprog'' solver in Optimization Toolbox of MATLAB 7.8.0. We have
used the options (MaxNodes, MaxRLPIter, and MaxTime) in the
``bintprog'' solver to greatly reduce the computational time.} is
also plotted serving as the performance upper bound.

In both suboptimal schemes, the transmission mode of each MS as well
as its assisting RS (if the cooperation-strategy is selected) are
pre-assigned. In specific, a MS is assigned to the direct
transmission mode if it is within the inner circle of the cell and
the cooperative transmission modes otherwise. When it is assigned
the cooperative transmission modes, if the large-scale fading of the
BS-RS link and the MS-RS link is about the same, two-way relaying is
adopted, otherwise one-way relaying is used (In this case, we assume
downlink is in direct transmission mode and uplink is in one-way
relaying transmission mode). For those MSs who need RS assistance,
we assign the nearest RS to each MS. Once the transmission modes and
assisting RSs for all MSs are determined, the subcarrier assignment
can be performed adaptively or randomly. In specific, for the
adaptive scheme, the optimization is formulated as a MWCP and solved
using ACO based algorithm according to instantaneous channel
conditions. For the random scheme, the subcarriers are arbitrarily
allocated in each time slot.

It is observed in Fig.~\ref{fig:throughput2} that the proposed
ACO-based joint adaptive algorithm performs very close to the upper
bound and the performance gap decreases as the transmit power
increases.
One also observes that it outperforms the two suboptimal schemes by
a significant margin. In particular, the tremendous improvement over
the suboptimal scheme with random subcarrier assignment clearly
demonstrates the benefits of multiuser diversity through adaptive
subcarrier assignment. The improvement over the suboptimal scheme
with adaptive subcarrier assignment further suggests the benefits of
transmission mode adaptation and relay selection.


%

The above simulation results in both Fig.~\ref{fig:throughput1} and
Fig.~\ref{fig:throughput2} show that the cooperative diversity gain,
network coding gain and multiuser diversity gain are efficiently
achieved by the proposed transmission protocol together with the
ACO-based resource allocation algorithm.

\subsection{Effect of Relay locations}
\begin{figure}[tbhp]
\begin{centering}
\includegraphics[scale=.8]{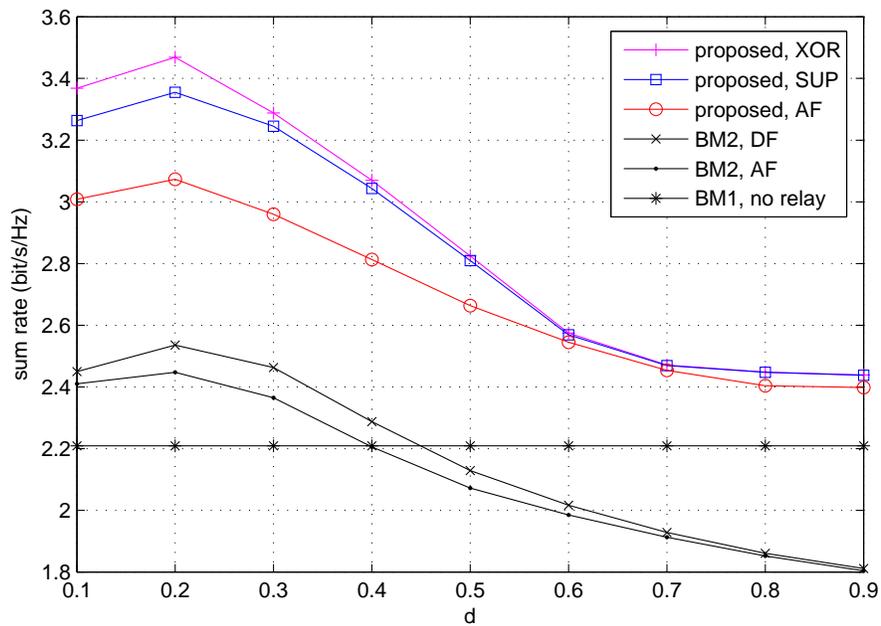}
\vspace{-0.1cm}
 \caption{Performance versus RS locations. $P_B=10$ dB per subcarrier. $d$ is the distance ratio of RSs inner circle radius to the cell radius.}\label{fig:relayvary}
\end{centering}
\vspace{-0.3cm}
\end{figure}

Finally, we investigate the impact of different relay locations on
the system throughput. In Fig.~\ref{fig:relayvary}, we fix
per-subcarrier BS power constraint $P_B=10$ dB, $d$ denotes the
distance ratio of the RSs located inner circle radius to the cell
radius. We can see that our proposed transmission protocol
outperforms BM1 and BM2 greatly, whatever $d$ varies. This further
illustrates the superiority of our proposed transmission protocol.
The maximum rate is achieved at about $d=0.2$ for all the considered
cooperation schemes. In addition, BM2 is inferior to BM1 when
$d>0.45$.
These observations show the relay location plays a key role in
achieving good performance in practical systems. In particular, our
results show that the RSs should be located closer to BS when MSs
are uniformly distributed in the cell.
By comparing the performance achieved by different relay strategies,
it is seen that whether to use DF or AF does not differ much in the
BM2. This conclusion is consistent with the previous work in
\cite{Meng05}. However, under the proposed transmission protocol, DF
is more favorable than AF.


\section{Conclusion}
In this paper, we proposed a novel three-time-slot TDD transmission
protocol for supporting direct transmission, one- and two-way
relaying in relay-assisted bidirectional cellular OFDMA networks.
Under this protocol, a complete set of five transmission modes are
introduced. We then formulated a combinatorial optimization problem
to jointly optimize subcarrier assignment, transmission mode
selection and relay selection for the system total throughput
maximization.
After establishing its equivalence to a maximum weighted clique
problem in graph theory, we employed an ACO based heuristic
algorithm to find the solution in polynomial time.
A few important conclusions have been made through extensive
simulations. Firstly, the proposed optimization framework can
achieve cooperative diversity gain, network coding gain and
multiuser diversity gain simultaneously and hence considerably
outperforms the existing benchmark schemes. In particular, about
20-30\% improvement on the system average throughput is achieved
over the conventional OFDMA networks without relay. Secondly,
choosing the appropriate transmission modes is necessary. Thirdly,
in a cell where MSs are uniformly distributed, it is better to place
the RSs closer to the BS rather in the middle of the cell.
Last but not least, DF relay strategy is practically better than AF
strategy under the proposed transmission protocol.

The proposed optimization framework can be extended if both user
fairness and asymmetric uplink and downlink traffic are taken into
account.


\appendices

\section{ACO Algorithm for MWCP}\label{app:aco}
\small{
\begin{tabular}{l}
  \hline
  \textbf{Main Function} \\ \hline
  1.  Initialize pheromone trails to $\tau_{max}$, $\mathcal {C}_{best}\leftarrow\emptyset$. \\
  2.  \textbf{repeat} \\
  3.  ~~~\textbf{for} each ant $a=1:nbAnts$, \textbf{do:} \\
  4.  ~~~~~~Construct clique $\mathcal {C}_{a}$. \\
  5.  ~~~\textbf{end for} \\
  6.  ~~~$\mathcal {C}_{iter}\leftarrow heaviest\{\mathcal {C}_1,...,\mathcal
  {C}_{nbAnts}\}$. \\
  7.  ~~~\textbf{if} $\mathcal {W}(\mathcal {C}_{iter})>\mathcal {W}(\mathcal {C}_{best})$, \textbf{do:} \\
  8.  ~~~~~~$\mathcal {C}_{best}\leftarrow\mathcal {C}_{iter}$. \\
  9.  ~~~\textbf{end if} \\
  10.~~~Update pheromone trails.  \\
  11. \textbf{until} the optimal solution is found or the maximum number of iterations reaches. \\
  12. \textbf{return} the largest weight constructed clique since the beginning. \\
  \hline  \\ \hline
  \textbf{Sub-Function} Construct clique \\ \hline
  1. Randomly choose a first vertex $v_f\in\mathcal {V}$. \\
  2. $\mathcal {C}\leftarrow\{v_f\}$. \\
  3. $Cadidates\leftarrow \{v_i|(v_f,v_i)\in\mathcal {E}\}$. \\
  4. \textbf{while} $Candidates\neq\emptyset$, \textbf{do:} \\
  5. ~~~Choose a vertex $v_i\in Candidates$ with probability $p(v_i)=\frac{[\tau_{\mathcal {C}}(v_i)]^\alpha}{\Sigma_{v_j\in Candidates}[\tau_{\mathcal {C}}(v_j)]^\alpha}$. \\
  6. ~~~$\mathcal {C}\leftarrow \mathcal {C}\cup\{v_i\}$. \\
  7. ~~~$Candidates\leftarrow Candidates\cap \{v_j|(v_i,v_j)\in\mathcal {E}\}$. \\
  8. \textbf{end while} \\
  9. return $\mathcal {C}$. \\ \hline \\ \hline
  \textbf{Sub-Function} Update pheromone trails \\ \hline
  1. \textbf{if} $v_i\in\mathcal {C}_{best}$, \textbf{do:} \\
  2. ~~~$\tau(v_i)\leftarrow \rho\tau(v_i)+1/\big(1+\mathcal {W}(\mathcal {C}_{best})-\mathcal {W}(\mathcal{C}_{iter})\big)$. \\
  3. \textbf{else} ~~\textbf{do:} \\
  4. ~~~$\tau(v_i)\leftarrow \rho\tau(v_i)$.\\
  5. \textbf{end if} \\
  6. \textbf{if} a pheromone trail is lower than $\tau_{min}$ \textbf{then} set it to $\tau_{max}$. \\
  7. \textbf{if} a pheromone trail is greater than $\tau_{max}$ \textbf{then} set it to
  $\tau_{min}$.\\ \hline
\end{tabular}}
%

\bibliographystyle{IEEEtran}
\bibliography{IEEEabrv,Yuan}


%
\begin{table}[htbp]
\renewcommand{\arraystretch}{1.3}
\caption{Transmission Mode Pairing} \label{pairing} \centering
\begin{tabular}{|c|c|c|c|}\hline
\backslashbox{downlink}{uplink} & {Direct} & {One-Way} &
{Two-Way}\\\hline Direct & {$\surd$} & {$\surd$} &
{$\times$}\\\hline One-Way & {$\surd$} & {$\surd$} &
{$\times$}\\\hline Two-Way & {$\times$} & {$\times$} &
{$\surd$}\\\hline
\end{tabular}
\end{table}

\begin{table}[htbp]
\caption{Graphical Settings} \label{tab:graph} \centering
\begin{tabular}{p{50pt}p{50pt}} \toprule
parameter & value \\
\midrule $\tau_{min}$ & 0.01 \\
$\tau_{max}$ & 6 \\
$\alpha$ & 1 \\
$\rho$ & 0.99 \\
ants & 10 \\
iterations & 500 \\ \bottomrule
\end{tabular}
\end{table}

%

%

%

\end{document}